\documentclass[english,notitlepage,nofootinbib]{revtex4-1}
\usepackage[T1]{fontenc}
\usepackage[latin9]{inputenc}
\usepackage{color}
\usepackage{amsmath}
\usepackage{amssymb}
\usepackage{graphicx}

\makeatletter

\providecommand{\tabularnewline}{\\}

\makeatother

\usepackage{babel}
\begin{document}

\title{Tree-level vacuum stability of two-Higgs-doublet models and new
constraints on the scalar potential}

\date{\today}

\author{Xun-Jie Xu }

\affiliation{Max-Planck-Institut f\"ur Kernphysik, Postfach 103980, D-69029 Heidelberg,
Germany.}
\begin{abstract}
The scalar potential of the two-Higgs-doublet model (2HDM)  may have
more than one local minimum and the usually considered vacuum could
be located at one  of them that could decay to another.  This
paper studies the condition that the usually considered vacuum is
the global minimum  which, combined with the bounded-from-below condition,
will stabilize the vacuum at tree level.  We further apply these
conditions to a specific 2HDM and obtain new constraints which could
 be important in phenomenological studies.
\end{abstract}
\maketitle

\section{Introduction}

As a simple extension of the Standard Model (SM), the Two-Higgs-Doublet
Model (2HDM) is well motivated in many  aspects, including supersymmetry
\cite{Djouadi:2005gj}, $CP$ violation \cite{Lee:1973iz}, axion
models \cite{Kim:1986ax},  etc. Besides, it also provides a very
rich phenomenology in collider experiments \cite{Haber:2015pua,Chakrabarty:2017qkh,CapdequiPeyranere:1990qk,Haber:2012vf,Ferreira:2012nv,Ferreira:2014naa,Ferreira:2014sld,Broggio:2014mna,Bernon:2015qea,Bernon:2015wef,Ferreira:2009jb,Ferreira:2011aa,Ferreira:2012my,Barroso:2013zxa,Ferreira:2013qua,Ferreira:2014dya,Ferreira:2014qda,Dumont:2014wha,Dev:2014yca,Campos:2017dgc}.
Therefore the 2HDM, as well as many variations, have been extensively
studied in recent years (see, e.g., \cite{Branco:2011iw} for a comprehensive
review).

In the 2HDM, an additional Higgs doublet is introduced to the scalar
sector of the SM. The scalar potential contains not only  self-coupling
terms of each Higgs doublet  but also several mixing terms of the
two doublets. As a consequence, the potential may contain several
different minima at which the scalar fields may obtain very different
vacuum expectation values (VEVs).  Depending on the configuration
of the potential, it is possible that one of the doublets does not
acquire a VEV (which appears in inert 2HDMs---see, e.g. \cite{Ma:2006km,Ma:2006fn,Barbieri:2006dq,Majumdar:2006nt,LopezHonorez:2006gr,Sahu:2007uh,Martinez:2011ua,Melfo:2011ie,Khan:2015ipa}),
or the VEVs break the $CP$ symmetry \cite{Dubinin:2002nx,Gunion:2005ja,Maniatis:2007vn,Bian:2016awe,Grzadkowski:2016szj},
or even break the electromagnetic $U(1)$ symmetry which should be
avoided in model building. In many phenomenological studies on 2HDMs,
the desired vacuum is usually  imposed  without checking whether
the potential necessarily results in such vacuum. It has been discussed
in  \cite{Ferreira:2004yd,Barroso:2005tq,Barroso:2005sm,Barroso:2005da,Ivanov:2006yq,Barroso:2006pa,Ivanov:2007de,Barroso:2007rr,Ivanov:2007ja,Ivanov:2008er,Ivanov:2010ww,BarroseSa:2009ak,Ginzburg:2009dp,Ivanov:2010wz,Battye:2011jj,Barroso:2012mj,Barroso:2013ica,Barroso:2013awa,Barroso:2013kqa,Ivanov:2015nea},
however, that more than one local minimum could coexist in the 2HDM
potential,  which implies the desired vacuum\footnote{In this paper we refer to the vacuum that is compatible with all experimental
observations as the desired vacuum.} might be only a local minimum that could decay into a deeper one
by quantum tunneling \cite{Coleman:1977py,Callan:1977pt}. If this
could happen, the vacuum would be unstable.

To avoid vacuum instability we expect a global minimum\footnote{In principle, if the vacuum is a local minimum but the decay rate
is low enough, then it can be metastable and  the model is still valid.
We will discuss this issue in Sec.\ \ref{sec:constraint}. } for the desired vacuum, which requires some new constraints on the
potential parameters, in addition to the bounded-from-below (BFB)
constraints \cite{Klimenko:1984qx,Maniatis:2006fs,Ivanov:2006yq,Ferreira:2009jb}.
Although it has been noticed in the literature \cite{Ferreira:2004yd,Barroso:2005tq,Barroso:2005sm,Barroso:2005da,Ivanov:2006yq,Barroso:2006pa,Ivanov:2007de,Barroso:2007rr,Ivanov:2007ja,Ivanov:2008er,Ivanov:2010ww,BarroseSa:2009ak,Ginzburg:2009dp,Ivanov:2010wz,Battye:2011jj,Barroso:2012mj,Barroso:2013ica,Barroso:2013awa,Barroso:2013kqa,Ivanov:2015nea}
that the vacuum could be unstable due to localness of the minimum,
in most phenomenological studies only the BFB constraint is taken
into account for the vacuum stability.\footnote{ In addition to the BFB constraint, the unitarity bound \cite{Casalbuoni:1986hy,Casalbuoni:1987eg,Maalampi:1991fb,Kanemura:1993hm,Arhrib:2000is,Akeroyd:2000wc,Ginzburg:2003fe,Ginzburg:2005dt,Horejsi:2005da}
is another theoretical constraint on the potential which has been
taken into account in many phenomenological studies.} Actually in the Higgs triplet model {[}the SM extended by an $SU(2)_{L}$
triplet Higgs{]}, recently we have derived explicit expressions of
the conditions to keep the desired vacuum globally minimal \cite{Xu:2016klg}.
It turns out that such conditions lead to  interesting constraints
on the masses of the scalar bosons in the Higgs triplet model. Therefore,
we expect that in 2HDMs this issue may also have phenomenologically
interesting consequences and should be taken into consideration in
future studies on 2HDMs. 

In this paper, we will investigate the 2HDM potential and derive the
condition for a selected minimum being the global one in order to
stabilize the corresponding vacuum. The method is similar to \cite{Xu:2016klg},
in which we first analytically compute all possible local minima and
then compare them with each other. We will adopt the most general
potential  including all renormalizable terms that respect the gauge
symmetries of the SM.  In some specific models, e.g. the type I and
type II 2HDMs with $\mathbb{Z}_{2}$ symmetries\footnote{The type I 2HDM couples all quarks to one Higgs doublet $\phi_{2}$
(denoting the other as $\phi_{1}$), which can be realized by the
$\mathbb{Z}_{2}$ symmetry: $\phi_{1}\rightarrow-\phi_{1}$, $\phi_{2}\rightarrow\phi_{2}$.
The type II 2HDM couples down-type right-handed quarks $d_{R}$ and
up-type right-handed quarks $u_{R}$ to $\phi_{1}$ and $\phi_{2}$
respectively, which can be realized by assigning the $\mathbb{Z}_{2}$
charge  to $d_{R}$ and $\phi_{1}$. In both cases,  terms in the
scalar potential containing odd numbers of $\phi_{1}$ should be absent.} to avoid tree-level flavour-changing neutral currents (FCNC), some
terms in the potential are absent.  They can be regarded as special
cases of the most general potential we adopted so our calculations
also apply to these special cases. Based on the analytical calculations,
a  numerical process is established to determine whether a selected
minimum is globally minimal. Applying this process to a specific $CP$-conserving
type I 2HDM with softly broken $\mathbb{Z}_{2}$ symmetry \cite{Haber:2015pua},
we show that the parameter space of the scalar potential can be considerably
constrained by the vacuum stability and in a certain scenario the
constraint is even stronger than the LHC constraint.

The issue that the 2HDM vacuum could be unstable if it was not the
global minimum has been studied in the literature for more than a
decade \cite{Ferreira:2004yd,Barroso:2005tq,Barroso:2005sm,Barroso:2005da,Ivanov:2006yq,Barroso:2006pa,Ivanov:2007de,Barroso:2007rr,Ivanov:2007ja,Ivanov:2008er,Ivanov:2010ww,BarroseSa:2009ak,Ginzburg:2009dp,Ivanov:2010wz,Battye:2011jj,Barroso:2012mj,Barroso:2013ica,Barroso:2013awa,Barroso:2013kqa,Ivanov:2015nea}.
In an early work \cite{Ferreira:2004yd} it has been shown that, for
a potential without explicit $CP$ breaking, if a minimum preserving
the electromagnetic $U(1)$ and $CP$ symmetries exists, then it is
the global one. However it was also pointed out in \cite{Barroso:2005sm,Barroso:2007rr}
that two neutral minima may coexist and have different potential depths.
References \cite{Ivanov:2006yq,Ivanov:2007de} adopt a geometric approach
by reformulating the 2HDM potential in terms of 3-quadrics in the
Minkowski space and prove that the potential can have at most two
local minima. Furthermore, if the two local minima coexist and there
is a discrete symmetry in the potential, then the two minima will
both break or both preserve the symmetry. In a recent study \cite{Ivanov:2015nea}
the criteria to guarantee that the desired minimum is global have
 been proposed. The method involves calculating a determinant and
in some cases solving eigenvalues of a $4\times4$ matrix numerically,
which is different from the method we adopt in this paper.

The reminder of this paper is organized as follows. In Sec.\ \ref{sec:local}
we analyze the most general scalar potential in 2HDMs and analytically
compute all possible types of local minima. Then we discuss how to
determine the global minimum in Sec.\ \ref{sec:glo} with a numerical
example to illustrate the method.  In Sec.\ \ref{sec:constraint}
we study the vacuum stability with both the BFB condition and the
requirement of a global minimum taken into account, focusing on the
type I 2HDM with softly broken $\mathbb{Z}_{2}$ symmetry. Finally,
we summarize at Sec.\ \ref{sec:Conclusion}. Some numerical examples
which can be used to verify our calculations are described in detail
in Appendix\ \ref{sec:Numerical}.

\section{The scalar potential and local minima\label{sec:local}}

With two Higgs doublets $\phi_{1}$ and $\phi_{2}$ (both have the
same hypercharge $Y=+1$), the most general renormalizable scalar
potential can be written as \cite{Wu:1994ja}
\begin{eqnarray}
V & = & m_{11}^{2}\phi_{1}^{\dagger}\phi_{1}+m_{22}^{2}\phi_{2}^{\dagger}\phi_{2}-\left[m_{12}^{2}\left(\phi_{1}^{\dagger}\phi_{2}\right)+{\rm h.c.}\right]\nonumber \\
 &  & +\frac{\lambda_{1}}{2}\left(\phi_{1}^{\dagger}\phi_{1}\right)^{2}+\frac{\lambda_{2}}{2}\left(\phi_{2}^{\dagger}\phi_{2}\right)^{2}+\lambda_{3}\left(\phi_{1}^{\dagger}\phi_{1}\right)\left(\phi_{2}^{\dagger}\phi_{2}\right)+\lambda_{4}\left|\phi_{1}^{\dagger}\phi_{2}\right|^{2}\nonumber \\
 &  & +\left[\frac{\lambda_{5}}{2}\left(\phi_{1}^{\dagger}\phi_{2}\right)^{2}+\lambda_{6}\left(\phi_{1}^{\dagger}\phi_{1}\right)\left(\phi_{1}^{\dagger}\phi_{2}\right)+\lambda_{7}\left(\phi_{2}^{\dagger}\phi_{2}\right)\left(\phi_{1}^{\dagger}\phi_{2}\right)+{\rm h.c.}\right].\label{eq:2h}
\end{eqnarray}
There are three quadratic terms and seven quartic terms in Eq.~(\ref{eq:2h}),
with four complex coefficients ($m_{12}^{2}$, $\lambda_{5,6,7}$)
and six real coefficients ($m_{11}^{2}$, $m_{22}^{2}$, $\lambda_{1,2,3,4}$),
i.e., 14 real parameters in total. However due to the unitary basis
transformation $(\phi_{1},\thinspace\phi_{2})^{T}\rightarrow U(\phi_{1},\thinspace\phi_{2})^{T}$
where $U$ is an $SU(2)$ matrix, three unphysical degrees of freedom
can be removed so actually there are only 11 physical degrees of freedom
\cite{Branco:2011iw}. In some 2HDMs due to additional symmetries
{[}e.g. $\mathbb{Z}_{2}$, $U(1)$, $U(2)$, etc.{]}, some terms are
absent. To apply the calculations  below to these specific models,
one only  needs to set the corresponding couplings to zero.  There
are 8 degrees of freedom in the two doublets $\phi_{1}$ and $\phi_{2}$.
After spontaneous symmetry breaking, three of them become Goldstone
bosons and the remaining five form massive scalar particles, including
two $CP$-even scalar bosons ($h$, $H$), one $CP$-odd scalar boson
($A$) and one charged scalar boson ($H^{\pm}$).

An important feature of the above potential is that it is a quadratic
functions of  three $SU(2)_{L}$ invariants of the fields \cite{Botella:1994cs}
\begin{equation}
q_{1}\equiv\phi_{1}^{\dagger}\phi_{1},\thinspace q_{2}\equiv\phi_{2}^{\dagger}\phi_{2},\thinspace z\equiv\phi_{1}^{\dagger}\phi_{2},\label{eq:2h-2}
\end{equation}
where $q_{1,2}$ are real, non-negative and $z$ is complex. Since
$\phi_{1}$ and $\phi_{2}$ are not invariant under $SU(2)_{L}$ transformations,
we will use $(q_{1},\thinspace q_{2},\thinspace z)$ instead of $(\phi_{1},\thinspace\phi_{2})$
in the following analysis. The potential expressed in terms of $q_{1,2}$
and $z$ is: 
\begin{eqnarray}
V & = & m_{11}^{2}q_{1}+m_{22}^{2}q_{2}-\left[m_{12}^{2}z+{\rm h.c.}\right]\nonumber \\
 &  & +\frac{\lambda_{1}}{2}q_{1}^{2}+\frac{\lambda_{2}}{2}q_{2}^{2}+\lambda_{3}q_{1}q_{2}+\lambda_{4}\left|z\right|^{2}\nonumber \\
 &  & +\left[\frac{\lambda_{5}}{2}z^{2}+\lambda_{6}q_{1}z+\lambda_{7}q_{2}z+{\rm h.c.}\right].\label{eq:2h-1}
\end{eqnarray}

Without boundary conditions, one can immediately find a minimum (to
distinguish it from other minima, we will refer to it as the type
A minimum) by solving 
\begin{equation}
\frac{\partial V}{\partial q_{1}}=\frac{\partial V}{\partial q_{2}}=\frac{\partial V}{\partial z}=0.\quad({\rm type\thinspace A})\label{eq:2h-3}
\end{equation}
Equation.~(\ref{eq:2h-3}) is a combination of linear equations with
respect to $(q_{1},\thinspace q_{2},\thinspace z)$, with the following
explicit form: 
\begin{equation}
\begin{cases}
m_{11}^{2}+q_{1}\lambda_{1}+q_{2}\lambda_{3}+z\lambda_{6}+\lambda_{6}^{*}z^{*} & =0\thinspace,\\
m_{22}^{2}+q_{2}\lambda_{2}+q_{1}\lambda_{3}+z\lambda_{7}+\lambda_{7}^{*}z^{*} & =0\thinspace,\\
-m_{12}^{2}+q_{1}\lambda_{6}+q_{2}\lambda_{7}+z\lambda_{5}+\lambda_{4}z^{*} & =0\thinspace.
\end{cases}\label{eq:2h-4}
\end{equation}
The solution of the above linear equation is given by \cite{Ferreira:2004yd,Maniatis:2006fs,Ivanov:2006yq,Nishi:2007nh}
\begin{equation}
\left(\begin{array}{c}
q_{1}\\
q_{2}\\
z\\
z^{*}
\end{array}\right)=\Lambda^{-1}b,\label{eq:2h-5}
\end{equation}
where\footnote{Note that in many models due to some symmetries in the potential,
the matrix $\Lambda$ could be noninvertible. For example, if $\lambda_{5,6,7}$
are real and $\lambda_{5}=\lambda_{4}$ then $\Lambda^{-1}$ does
not exist. However, one can still use Eqs.~(\ref{eq:2h-5}) and ~(\ref{eq:2h-6})
in this case by taking $\Lambda^{-1}\equiv\lim_{\delta\Lambda\rightarrow0}(\Lambda+\delta\Lambda)^{-1}$.
In Sec. \ref{sec:glo} we will present an example in this case.}
\begin{equation}
\Lambda\equiv\left(\begin{array}{cccc}
\lambda_{1} & \lambda_{3} & \lambda_{6} & \lambda_{6}^{*}\\
\lambda_{3} & \lambda_{2} & \lambda_{7} & \lambda_{7}^{*}\\
\lambda_{6} & \lambda_{7} & \lambda_{5} & \lambda_{4}\\
\lambda_{6}^{*} & \lambda_{7}^{*} & \lambda_{4} & \lambda_{5}^{*}
\end{array}\right),\thinspace b\equiv\left(\begin{array}{c}
-m_{11}^{2}\\
-m_{22}^{2}\\
m_{12}^{2}\\
m_{12}^{*2}
\end{array}\right).\label{eq:2h-6}
\end{equation}
The potential value at this minimum is 
\begin{equation}
V_{{\rm min,\thinspace A}}=-\frac{1}{2}b^{T}\Lambda^{-1}b.\label{eq:2h-7}
\end{equation}

However, we should note that the three variables $(q_{1},\thinspace q_{2},\thinspace z)$
are subjected to some boundary conditions. From the definitions of
$q_{1}$ and $q_{2}$ we have two boundary conditions
\begin{eqnarray}
({\rm i}): & \quad & q_{1}\geq0;\label{eq:2h-8}\\
({\rm ii}): & \quad & q_{2}\geq0.\label{eq:2h-9}
\end{eqnarray}
Besides, since $z$ is a scalar product of two complex vectors, its
absolute value should be less than or equal to the product of their
lengths, which is $\sqrt{q_{1}}\sqrt{q_{2}}$. So we have
\begin{equation}
({\rm iii}):\quad q_{1}q_{2}\geq|z|^{2}.\label{eq:2h-10}
\end{equation}
The type A minimum exists only if the point computed via Eq.\ (\ref{eq:2h-5})
is located in the region restricted by the above conditions.

Apart from the type A minimum, other minima could exist but they should
be on the boundaries (\ref{eq:2h-8}), (\ref{eq:2h-9}) or (\ref{eq:2h-10})
otherwise they would be determined by the off-boundary minimization
equation (\ref{eq:2h-3}) which has a unique solution, i.e. the type
A minimum. There are four types of on-boundary minima, which depending
on the boundaries will be referred to as the type B, C, D and E minima:
\begin{eqnarray}
{\rm type\thinspace B}: & \quad & q_{1}=0,\thinspace q_{2}>0,\thinspace z=0;\label{eq:2h-11}\\
{\rm type\thinspace C}: & \quad & q_{2}=0,\thinspace q_{1}>0,\thinspace z=0;\label{eq:2h-12}\\
{\rm type\thinspace D}: & \quad & q_{1}>0,\thinspace q_{2}>0,\thinspace q_{1}q_{2}=|z|^{2};\label{eq:2h-13}\\
{\rm type\thinspace E}: & \quad & q_{1}=0,\thinspace q_{2}=0,\thinspace z=0.\label{eq:2h-14}
\end{eqnarray}

Type B and C minima can be computed by setting $q_{1}$ or $q_{2}$
to zero and minimizing the potential with respect to $q_{2}$ or $q_{1}$.
This gives
\begin{eqnarray}
{\rm type\thinspace B}: & \quad & q_{1}=0,\thinspace q_{2}=-\frac{m_{22}^{2}}{\lambda_{2}},\thinspace z=0;\label{eq:2h-11-1}\\
{\rm type\thinspace C}: & \quad & q_{2}=0,\thinspace q_{1}=-\frac{m_{11}^{2}}{\lambda_{1}},\thinspace z=0.\label{eq:2h-12-1}
\end{eqnarray}
The corresponding potential values at these minima are given by
\begin{equation}
V_{{\rm min,\thinspace B}}=-\frac{m_{22}^{4}}{2\lambda_{2}},\label{eq:2h-7-1}
\end{equation}
\begin{equation}
V_{{\rm min,\thinspace C}}=-\frac{m_{11}^{4}}{2\lambda_{1}}.\label{eq:2h-7-1-1}
\end{equation}

Type E is the simplest case. All fields and the potential value are
zero at this minimum. 

The remaining case, type D, is actually the desired vacuum in many
2HDMs with non-vanishing $\langle\phi_{1}\rangle$ and $\langle\phi_{2}\rangle$,
because $q_{1}q_{2}=|z|^{2}$ implies the two complex vectors $\phi_{1}$
and $\phi_{2}$ are parallel to each other  at the minimum, i.e. $\langle\phi_{1}\rangle\propto\langle\phi_{2}\rangle$.
Since the absolute value of $z$ is fixed by $q_{1}q_{2}$, the potential
can be treated as a function of $q_{1}$, $q_{2}$ and
\begin{equation}
\theta\equiv\arg z.\label{eq:2h-20}
\end{equation}
From $\partial V/\partial q_{1}=\partial V/\partial q_{2}=\partial V/\partial\theta=0$
we get
\begin{eqnarray}
m_{11}^{2} & = & m_{12R}^{2}t_{\beta}-q\left[\lambda_{1}c_{\beta}^{2}+(\lambda_{3}+\lambda_{4}+\lambda_{5R})s_{\beta}^{2}+3\lambda_{6R}c_{\beta}s_{\beta}+\lambda_{7R}t_{\beta}s_{\beta}^{2}\right],\label{eq:2h-23}\\
m_{22}^{2} & = & m_{12R}^{2}t_{\beta}^{-1}-q\left[\lambda_{2}s_{\beta}^{2}+(\lambda_{3}+\lambda_{4}+\lambda_{5R})c_{\beta}^{2}+\lambda_{6R}t_{\beta}^{-1}c_{\beta}^{2}+3\lambda_{7R}c_{\beta}s_{\beta}\right],\label{eq:2h-24}\\
m_{12I}^{2} & = & q\left[\lambda_{5I}c_{\beta}s_{\beta}+\lambda_{6I}c_{\beta}^{2}+\lambda_{7I}s_{\beta}^{2}\right],\label{eq:2h-25}
\end{eqnarray}
where we have defined
\begin{equation}
q\equiv q_{1}+q_{2},\label{eq:2h-19-1}
\end{equation}
\begin{equation}
\tan^{2}\beta\equiv\frac{q_{2}}{q_{1}},\quad\beta\in[0,\frac{\pi}{2}],\label{eq:2h-16-1}
\end{equation}
\begin{equation}
(m_{12R}^{2},\thinspace\lambda_{5R},\thinspace\lambda_{6R},\thinspace\lambda_{7R})\equiv{\rm Re}(m_{12}^{2}e^{i\theta},\thinspace\lambda_{5}e^{2i\theta},\thinspace\lambda_{6}e^{i\theta},\thinspace\lambda_{7}e^{i\theta}),\label{eq:2h-21}
\end{equation}
\begin{equation}
(m_{12I}^{2},\thinspace\lambda_{5I},\thinspace\lambda_{6I},\thinspace\lambda_{7I})\equiv{\rm Im}(m_{12}^{2}e^{i\theta},\thinspace\lambda_{5}e^{2i\theta},\thinspace\lambda_{6}e^{i\theta},\thinspace\lambda_{7}e^{i\theta}).\label{eq:2h-22}
\end{equation}

The above calculation is basis independent. However, to represent
it in a more conventional form, we may choose an appropriate basis
so that
\begin{equation}
\langle\phi_{1}\rangle=\frac{1}{\sqrt{2}}\left(\begin{array}{c}
0\\
v_{1}
\end{array}\right),\thinspace\langle\phi_{2}\rangle=\frac{1}{\sqrt{2}}\left(\begin{array}{c}
0\\
v_{2}e^{i\theta}
\end{array}\right),\label{eq:2h-15}
\end{equation}
which implies that the type D minimum corresponds to  the vacuum usually
adopted in many 2HDMs. In this basis, one can interpret $\tan\beta$
as the well-known ratio
\begin{equation}
\tan\beta=\frac{v_{2}}{v_{1}},\label{eq:2h-18}
\end{equation}
and relate the value of $q$ at the minimum to the electroweak vacuum
expectation value
\begin{equation}
v=\sqrt{v_{1}^{2}+v_{2}^{2}}=\sqrt{2q}\approx246\ {\rm GeV}.\label{eq:2h-26}
\end{equation}

One can solve Eqs.\ (\ref{eq:2h-23}),\ (\ref{eq:2h-24}), and\ (\ref{eq:2h-25})
with respect to $q$ (or $v^{2}$), $\beta$, and $\theta$ to get
the  type D minimum of the potential, at least numerically. Since
they are nonlinear equations, the solutions may be not unique. In
general, all solutions should be taken into account, which makes the
type D minima a little more complicated than the other types. We will
show an example which has more than one type D minimum in Sec.\ \ref{sec:glo}.

\begin{table}
\centering

\caption{\label{tab:All-local-minima}All possible local minima of the scalar
potential. ``$\times$''  denotes a nonzero component, and ``$*$''
stands for an arbitrary value (can be zero or nonzero). }

\begin{ruledtabular}

\begin{tabular}{cccccc}
 & \textcolor{white}{.}\hspace{0.8cm}Type A\hspace{0.8cm}\textcolor{white}{.} & \textcolor{white}{.}\hspace{0.8cm}Type B\hspace{0.8cm}\textcolor{white}{.} & \textcolor{white}{.}\hspace{0.8cm}Type C\hspace{0.8cm}\textcolor{white}{.} & \textcolor{white}{.}\hspace{0.8cm}Type D\hspace{0.8cm}\textcolor{white}{.} & \textcolor{white}{.}\hspace{0.8cm}Type E\hspace{0.8cm}\textcolor{white}{.}\tabularnewline
\hline 
\rule[-5ex]{0pt}{10ex}  $\langle\phi_{1}\rangle$, $\langle\phi_{2}\rangle$ & $\left[\begin{array}{c}
0\\
\times
\end{array}\right]$, $\left[\begin{array}{c}
\times\\
*
\end{array}\right]$ & $\left[\begin{array}{c}
0\\
0
\end{array}\right]$, $\left[\begin{array}{c}
0\\
\times
\end{array}\right]$ & $\left[\begin{array}{c}
0\\
\times
\end{array}\right]$, $\left[\begin{array}{c}
0\\
0
\end{array}\right]$ & $\left[\begin{array}{c}
0\\
\times
\end{array}\right]$, $\left[\begin{array}{c}
0\\
\times
\end{array}\right]$ & $\left[\begin{array}{c}
0\\
0
\end{array}\right]$, $\left[\begin{array}{c}
0\\
0
\end{array}\right]$\tabularnewline
\rule[-3ex]{0pt}{8ex}  $(q_{1},\thinspace q_{2},\thinspace z)$ & Eq.\ (\ref{eq:2h-5}) & $(0,\thinspace-\frac{m_{22}^{2}}{\lambda_{2}},\thinspace0)$ & $(-\frac{m_{11}^{2}}{\lambda_{1}},\thinspace0,\thinspace0)$ & Eqs.\ (\ref{eq:2h-23}),\ (\ref{eq:2h-24}),\ (\ref{eq:2h-25}) & $(0,\thinspace0,\thinspace0)$\tabularnewline
\rule[-5ex]{0pt}{10ex}  $\begin{array}{c}
{\rm Existence}\\
{\rm condition}
\end{array}$ & $\begin{array}{c}
q_{1},\thinspace q_{2}>0\\
|z|^{2}<q_{1}q_{2}
\end{array}$ & $q_{2}>0$ & $q_{1}>0$ & $\begin{array}{c}
q_{1},\thinspace q_{2}>0\\
|z|^{2}=q_{1}q_{2}
\end{array}$ & $/$ \footnote{Not required.}\tabularnewline
\hline 
\rule[-3ex]{0pt}{8ex}  $V_{{\rm min}}$ & Eq.\ (\ref{eq:2h-7}) & $-\frac{m_{22}^{4}}{2\lambda_{2}}$ & $-\frac{m_{11}^{4}}{2\lambda_{1}}$ & $/$ \footnote{Not unique.} & 0\tabularnewline
\end{tabular}

\end{ruledtabular}
\end{table}

Finally, let us summarize all possible local minima of the scalar
potential, as  listed in Tab. \ref{tab:All-local-minima}. There are
five types of minima, classified according to whether they locate
on some boundaries {[}cf. Eqs.~(\ref{eq:2h-8}),~(\ref{eq:2h-9}),
and (\ref{eq:2h-10}){]} in the $(q_{1},\thinspace q_{2},\thinspace z)$
space. Type A is not on any of the boundaries,  which implies it has
the most degrees of freedom in the minimization, while type E is actually
on all the boundaries so it is completely fixed by these boundaries.
To provide a straightforward understanding of them, the second row
of Tab. \ref{tab:All-local-minima} shows the zero components of $\langle\phi_{1}\rangle$
and $\langle\phi_{2}\rangle$, where we use ``$\times$'' and ``$*$''
to represent nonzero and arbitrary (zero or nonzero) values, respectively.
However, one should be careful about the basis dependence. For any
minimum of the potential, the transformation
\begin{equation}
\phi_{1}\rightarrow U\phi_{1},\thinspace\phi_{2}\rightarrow U\phi_{2},\label{eq:2h-27}
\end{equation}
where $U$ is a $2\times2$ unitary matrix would transform the vacuum
to another equivalent vacuum, while the appearance of zero components
in $\langle\phi_{1}\rangle$ and $\langle\phi_{2}\rangle$ is not
invariant under this  transformation. Therefore, $\langle\phi_{1}\rangle$
and $\langle\phi_{2}\rangle$ in Tab. \ref{tab:All-local-minima}
for each type of minima should be understood as a category of VEVs
that can be transformed to such a form. For instance, if the potential
is found to has a minimum at
\begin{equation}
\langle\phi_{1}\rangle=\left[\begin{array}{c}
\times\\
\times
\end{array}\right],\langle\phi_{2}\rangle=\left[\begin{array}{c}
\times\\
\times
\end{array}\right],\label{eq:2h-28}
\end{equation}
or
\begin{equation}
\langle\phi_{1}\rangle=\left[\begin{array}{c}
\times\\
\times
\end{array}\right],\langle\phi_{2}\rangle=\left[\begin{array}{c}
0\\
\times
\end{array}\right],\label{eq:2h-29}
\end{equation}
then this minimum should be of type A (generally), since the doublets
can be transformed to the form $U\langle\phi_{1}\rangle=(0,\thinspace\times)^{T}$
and $U\langle\phi_{2}\rangle=(\times,\thinspace*)^{T}$.  However,
if it happens coincidentally  that $\langle\phi_{1}\rangle\propto\langle\phi_{2}\rangle$
in Eq.~(\ref{eq:2h-28}), then this falls into type D, since the transformation
will simultaneously set the upper components of $\langle\phi_{1}\rangle$
and $\langle\phi_{2}\rangle$ to zero.

To avoid the basis dependence, we recommend using $(q_{1},\thinspace q_{2},\thinspace z)$
instead of explicit forms of $\langle\phi_{1}\rangle$ and $\langle\phi_{2}\rangle$.
Once a minimum is found, one can compute the values of $(q_{1},\thinspace q_{2},\thinspace z)$
to see whether the minimum is located on some of the boundaries of
Eqs.~(\ref{eq:2h-8}),~(\ref{eq:2h-9}) , and (\ref{eq:2h-10}), which
is the original definition of the five types of minima.

For any given potential in 2HDMs, one can exhaustively find all the
possible minima by computing $(q_{1},\thinspace q_{2},\thinspace z)$
according to the third row of Tab. \ref{tab:All-local-minima}. But
the corresponding minima do not necessarily exist, e.g. $q_{1}$ or
$q_{2}$ computed in this way may be negative. So in Tab. \ref{tab:All-local-minima}
we also provide the conditions of existence of these minima, which
should be checked after $(q_{1},\thinspace q_{2},\thinspace z)$ is
computed. The existence conditions listed here are only necessary
conditions, not sufficient, which implies the locations could be saddle
points or even local maxima. But in the framework of this paper, this
does not concern us because by comparing the potential values at these
points we will take the lowest point among these candidates as the
vacuum of the model, which must be the global minimum, provided that
the potential is BFB.

\section{Global minima\label{sec:glo}}

As we have derived, there can be several types of minima in the scalar
potential so it is possible that the desired vacuum  is not located
at a  global minimum. This problem in 2HDMs has been studied in
Refs.\ \cite{Ferreira:2004yd,Barroso:2005sm,Barroso:2007rr,Ivanov:2007de}
where one can find some useful conclusions. First, it has been proved
that at most two physically inequivalent local  minima can coexist
in the potential. This implies that among the five possible types
of local minima, only one or two of them can actually exist in a certain
potential while the others  should be saddle points or local maxima.
Second, the two minima (if exist) violate or conserve a discrete symmetry
(if exist in the potential) simultaneously, e.g., $CP$-conserving
and $CP$-violating minima cannot coexist in a $CP$-conserving potential. 

Despite these conclusions, given a general 2HDM potential and a minimum
of the potential, one still cannot determine the globalness of the
minimum in a simple way. However, since we know how to find all possible
local minima in the potential, there is a numerical process that enables
us to determine whether a minimum is global or local.  

For a given potential and one of its minima, denoting the location
of this minimum as $P$ and the corresponding potential value as $V_{P}$,
if $V_{P}>0$ then obviously $P$ is not a global minimum because
the potential value is larger than the type E minimum. If $V_{P}\leq0$,
then one proceeds as follows:
\begin{enumerate}
\item Compute $(q_{1},\thinspace q_{2},\thinspace z)$ for the minima of
types A, B, and C listed in Tab. \ref{tab:All-local-minima}.
\item Check the existence condition for each of them. If it is violated
then it will not be considered anymore.
\item Compute the potential values at the remaining minima; if any of them
are lower than $V_{P}$ then $P$ is not a global minimum.
\item Otherwise, numerically solve Eqs.\ (\ref{eq:2h-23}),\ (\ref{eq:2h-24})
and (\ref{eq:2h-25}) and compare $V_{P}$ with the potential values
of these solutions. If $V_{P}$ is still the lowest, then $P$ is
the global minimum.
\end{enumerate}
Here, in principle, we can also include type D in step 1 and remove
step 4. However, type D involves solving non-linear equations numerically,
which consumes much more CPU usage for computers to work it out than
types A, B and C. According to our numerical experience, if $P$ is
not a global minimum, in many cases it can be excluded by comparing
with the first three types of minima. Therefore we leave the calculation
of type D minima as the last step, which optimizes the program considerably. 

The above process is simple to be realized in programming so it can
be readily included in numerical analyses of 2HDMs. Actually, the
situation is similar to the BFB condition. Despite that there have
been some analytical expressions of the BFB condition in some special
cases such as $\lambda_{6}=\lambda_{7}=0$ \cite{Branco:2011iw},
for the most general potential there has not been a simple criterion
for BFB. Currently a numerical process that combines necessary analytical
results is able to achieve this, which has been adopted in the program
{\tt 2HDMC} \cite{Eriksson:2009ws}. Likewise, one may also adopt
the similar numerical process proposed above to check whether a minimum
is global.

To illustrate the above process, we will analyze a specific example,
 which also serves as a benchmark to show that requiring the vacuum
to be globally minimal provides new constraints on the model. In a
specific scenario of 2HDM studied in \cite{Haber:2015pua}, the potential
parameters take the following values:\footnote{Corresponding to $m_{h}=125$ GeV, $m_{H}=500$ GeV, $c_{\beta-\alpha}=0.1$,
$Z_{4}=-2$, $Z_{5}=-2$, $Z_{7}=0$ and $\tan\beta=30$ in the hybrid
basis \cite{Haber:2015pua}, which are taken as an  input in the code
{\tt 2HDMC} \cite{Eriksson:2009ws} to generate the parameters in
Eqs.~(\ref{eq:2h-30}) and (\ref{eq:2h-31}). }
\begin{equation}
(m_{11}^{2},m_{22}^{2},m_{12}^{2})=(-0.110625,-0.00831996,0.00827196)\ {\rm TeV}^{2},\label{eq:2h-30}
\end{equation}
\begin{equation}
\lambda_{1\cdots7}=(11.8234,0.270735,15.8106,-1.98716,-1.98716,0,0).\label{eq:2h-31}
\end{equation}
This example is well compatible with recent constraints from LHC \cite{Dumont:2014wha,Haber:2015pua},
including both the observed Higgs signal and non-observation of additional
Higgs states. The vacuum of this model is at $v=246.2$ GeV and $\tan\beta=30$,
corresponding to a minimum at 
\begin{equation}
P=(q_{1},\thinspace q_{2},\thinspace z)=(3.36\times10^{-5},\thinspace3.03\times10^{-2},\thinspace1.01\times10^{-3})\ {\rm TeV}^{2}.\label{eq:2h-61}
\end{equation}
The potential value at this minimum is  $V_{P}=-1.36\times10^{-4}$
${\rm TeV}^{4}$. 

Now we would like to know whether this minimum is global or local.
Since $V_{P}<0$, we neglect the type E minimum. Computing $(q_{1},\thinspace q_{2},\thinspace z)$
for type A, B and C minima gives\footnote{In this example we have $\lambda_{4}=\lambda_{5}$ which makes $\Lambda$
non-invertible. Such cases appear occasionally in 2HDM potentials
with some symmetries. To validate Eqs.~(\ref{eq:2h-5}) and ~(\ref{eq:2h-6}),
one can simply add a small perturbation $\delta\lambda_{5}$ to $\lambda_{5}$
and compute $\lim_{\delta\lambda_{5}\rightarrow0}\Lambda^{-1}$. }
\begin{equation}
\frac{(q_{1},\thinspace q_{2},\thinspace z)}{10^{-3}{\rm TeV}^{2}}=\begin{cases}
(0.412,\thinspace6.69,-2.08) & ({\rm type\thinspace A})\\
(0,\quad\quad30.7,\quad\quad0) & ({\rm type\thinspace B})\\
(9.36,\quad0,\quad\quad\quad0) & ({\rm type\thinspace C})
\end{cases}\label{eq:2h-32}
\end{equation}
However, the solution of type A violates its existence condition since
$|z|^{2}>q_{1}q_{2}$. Thus we only need to compute the potential
values at type B and C minima:
\begin{equation}
(V_{{\rm B}},\thinspace V_{{\rm {\rm C}}})=(-1.28,-5.18)\cdot10^{-4}\ {\rm TeV}^{4}.\label{eq:2h-33}
\end{equation}
As we can see, $V_{{\rm {\rm C}}}<V_{P}$ so we can conclude that
$P$ is not the global minimum without solving the nonlinear equations
of type D. 

Nevertheless,   if we further solve the equations of type D, we
will know the actual vacuum structure. According to the results  in
Appendix \ref{sec:Numerical}, there are four solutions of type D
in this example,  denoted as D1, D2, D3 and D4, with the following
potential values:
\begin{equation}
(V_{{\rm D1}},\thinspace V_{{\rm D2}},\thinspace V_{{\rm D3}},\thinspace V_{{\rm D4}})=(-5.24,-1.36,-1.04,-0.343)\times10^{-4}\ {\rm TeV}^{4}.\label{eq:2h-34}
\end{equation}
These values imply that  the global minimum of this potential is D1
while $P$ is actually identical to D2 which is a local minimum\footnote{In general, one needs to compute the second-order derivatives (Hessian
matrix) to make sure that it is not a saddle point or a local maximum.
However, this is not necessary for $P$ since we know the mass spectrum
of the scalar bosons (see Appendix \ref{sec:Numerical}) is above
zero, which implies the eigenvalues of the Hessian matrix are all
positive.}. Since there can be at most two physically inequivalent minima in
a 2HDM potential, we immediately know that these two minima must be
D1 and D2. As a consequence, the other local minimum candidates such
as $V_{{\rm B}}$ and $V_{{\rm {\rm C}}}$ in Eq.~(\ref{eq:2h-33})
can only be saddle points or local maxima, despite that $V_{{\rm {\rm C}}}$
is lower than the local minimum $V_{P}$.

\section{\label{sec:constraint}Tree-level vacuum stability and its constraints
on 2HDMs}

In the previous section, we have seen an example of the 2HDM potential
in which the desired vacuum is not a global minimum, though it is
compatible with all constraints from LHC. In such case, the vacuum
may decay into a deeper minimum via quantum tunneling which makes
the vacuum unstable. This situation should be avoided in any valid
2HDM. Consequently we will have a new constraint on the model by requiring
that the desired vacuum is a global minimum.

We would like to comment here that this  is not completely equivalent
to the vacuum stability. First, if the decay rate is low enough, the
vacuum can be metastable which means the lifetime of the vacuum is
longer than the age of the universe so that the metastable vacuum
still survives today. Second, even if the vacuum at tree level is
absolutely stable, in the effective potential including loop corrections
\cite{Staub:2017ktc,Dorsch:2017nza} the vacuum could be   unstable.
A well-known example is the SM in which the potential would be negative
around ${\cal O}(10^{10})$ GeV due to large loop corrections from
the top quark \cite{Lindner:1988ww}. An elaborate investigation \cite{Buttazzo:2013uya}
suggests that with current best-fit values of top quark and Higgs
masses the SM vacuum is metastable. 

In principle, one may take the metastability and loop corrections
into consideration in 2HDMs as well. However, both of them are only
important when the parameters of the model are near some critical
configurations (e.g. the SM metastability depends critically on the
top quark mass). In 2HDMs many parameters are still far from being
precisely determined while in contrast, the potential parameters of
the SM have been known precisely from the Higgs mass and the Fermi
constant. Therefore at the current stage, we should not be concerned
with these two issues. Instead, we only consider  the absolute stability
of vacuum at tree level.

\begin{figure}
\centering

\includegraphics{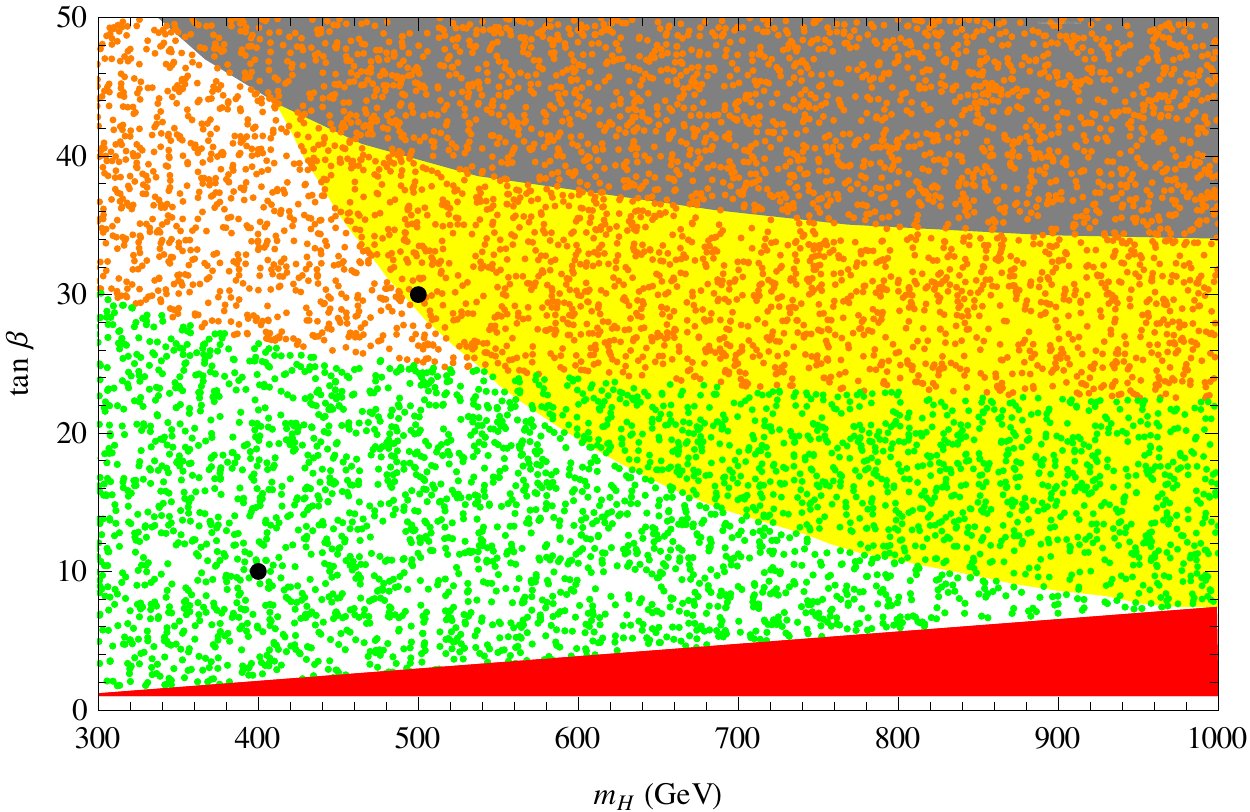}

\caption{\label{fig:Constraints}Constraints on $(m_{H},\thinspace\tan\beta)$
from absolute stability of the vacuum at tree-level. The green and
orange points denote samples of which the vacua are global and local
minima respectively.  The red region and the yellow region violate
the BFB and the unitarity bounds respectively. The gray region is
disfavored by direct searches from LHC for the type I 2HDM \cite{Haber:2015pua}.
The black points are two examples with numerical details presented
 in Appendix \ref{sec:Numerical}.}
\end{figure}

Next we will focus on a specific 2HDM to study the constraint from
the  vacuum stability on some physical quantities. In a $CP$-conserving
2HDM with a softly broken $\mathbb{Z}_{2}$ symmetry that has been
studied in Ref.\ \cite{Haber:2015pua}, the quartic terms are invariant
under the $\mathbb{Z}_{2}$ transformation 
\begin{equation}
(\phi_{1},\thinspace\phi_{2})\rightarrow(\phi_{1},-\phi_{2}),\label{eq:2h-35}
\end{equation}
which leads to
\begin{equation}
\lambda_{6}=\lambda_{7}=0.\label{eq:2h-36}
\end{equation}
The quadratic term $m_{12}^{2}\phi_{1}^{\dagger}\phi_{2}$, however,
softly breaks the $\mathbb{Z}_{2}$ symmetry. All quartic and quadratic
coefficients in the potential are real due to the $CP$ symmetry.
After spontaneous symmetry breaking, the two Higgs doublets are expected
to acquire $CP$-conserving VEVs:
\begin{equation}
\langle\phi_{1}\rangle=\frac{1}{\sqrt{2}}\left(\begin{array}{c}
0\\
v_{1}
\end{array}\right),\thinspace\langle\phi_{2}\rangle=\frac{1}{\sqrt{2}}\left(\begin{array}{c}
0\\
v_{2}
\end{array}\right).\label{eq:2h-37}
\end{equation}

Apart from three Goldstone bosons, there are four mass eigenstates
of the scalar fields, including two $CP$-even Higgs fields $h$ and
$H$, a $CP$-odd field $A$ and a charged Higgs $H^{\pm}$. Their
masses are given by \cite{Gunion:2002zf}
\begin{equation}
m_{A}^{2}=\frac{m_{12}^{2}}{s_{\beta}c_{\beta}}-\lambda_{5}v^{2},\label{eq:2h-38}
\end{equation}
\begin{equation}
m_{H^{\pm}}^{2}=m_{A}^{2}+\frac{1}{2}(\lambda_{5}-\lambda_{4})v^{2},\label{eq:2h-38-1}
\end{equation}
\begin{equation}
m_{H,h}^{2}=\frac{1}{2}[M_{11}^{2}+M_{22}^{2}\pm\sqrt{(M_{11}^{2}-M_{22}^{2})^{2}+4(M_{12}^{2})^{2}}],\label{eq:2h-39}
\end{equation}
where the matrix $M^{2}$ is defined as 
\[
M^{2}\equiv m_{A}^{2}\left(\begin{array}{cc}
s_{\beta}^{2} & -c_{\beta}s_{\beta}\\
-c_{\beta}s_{\beta} & c_{\beta}^{2}
\end{array}\right)+v^{2}\left(\begin{array}{cc}
c_{\beta}^{2}\lambda_{1}+s_{\beta}^{2}\lambda_{5} & c_{\beta}s_{\beta}\left(\lambda_{3}+\lambda_{4}\right)\\
c_{\beta}s_{\beta}\left(\lambda_{3}+\lambda_{4}\right) & s_{\beta}^{2}\lambda_{2}+c_{\beta}^{2}\lambda_{5}
\end{array}\right).
\]

Diagonalizing the matrix $M^{2}$ gives the eigenvalues $m_{H,h}^{2}$
and the mixing angle $\alpha$ ($-\pi/2$$\leq\alpha\leq$$\pi/2$)
between the two eigenstates,
\begin{equation}
\alpha=\frac{1}{2}\arg(M_{11}^{2}-M_{22}^{2}+2iM_{12}^{2}),\thinspace\alpha\in(-\frac{\pi}{2},\frac{\pi}{2}].\label{eq:2h-40}
\end{equation}

In Ref.\ \cite{Haber:2015pua}, a ``hybrid'' basis is adopted in
specifying the input parameters of the model. The input parameters
in the hybrid basis are $(m_{h},\thinspace m_{H},\thinspace c_{\beta-\alpha},\thinspace\tan\beta,\thinspace Z_{4},\thinspace Z_{5},\thinspace Z_{7})$
where $c_{\beta-\alpha}\equiv\cos(\beta-\alpha)$ and $Z_{4,5,7}$
are defined as
\begin{equation}
Z_{4}\equiv\frac{1}{4}s_{2\beta}^{2}\left[\lambda_{1}+\lambda_{2}-2\lambda_{345}\right]+\lambda_{4},\label{eq:2h-41}
\end{equation}
\begin{equation}
Z_{5}\equiv\frac{1}{4}s_{2\beta}^{2}\left[\lambda_{1}+\lambda_{2}-2\lambda_{345}\right]+\lambda_{5},\label{eq:2h-42}
\end{equation}
\begin{equation}
Z_{7}\equiv-\frac{1}{2}s_{2\beta}\left[\lambda_{1}s_{\beta}^{2}-\lambda_{2}c_{\beta}^{2}+\lambda_{345}c_{2\beta}\right].\label{eq:2h-43}
\end{equation}

This basis has already been included in the code {\tt 2HDMC} \cite{Eriksson:2009ws}
which facilitates the input of valid model parameters. Hence we will
use {\tt 2HDMC} in scanning the parameter space. We focus on a specific
scenario with the major portion of its parameter space still compatible
with the recent LHC constraints. In the hybrid basis, the input parameters
are 
\begin{equation}
(m_{h},\thinspace c_{\beta-\alpha},\thinspace Z_{4},\thinspace Z_{5},\thinspace Z_{7})=(125\thinspace{\rm GeV},\thinspace0.1,\thinspace-2,\thinspace-2,\thinspace0),\label{eq:2h-45}
\end{equation}
and 
\begin{equation}
300\thinspace{\rm GeV}\leq m_{H}\leq1000\thinspace{\rm GeV},\quad1\leq\tan\beta\leq50.\label{eq:2h-44}
\end{equation}

To show the constraint from the vacuum stability, we randomly generate
$10^{4}$ samples with the input parameters given in Eqs.~(\ref{eq:2h-45})
and (\ref{eq:2h-44}). We use {\tt 2HDMC} to compute the corresponding
potential parameters $(m_{11}^{2},m_{22}^{2},m_{12}^{2})$ and $\lambda_{1\cdots7}$
and also to check the BFB condition of the potentials. For those
samples with BFB potentials, we proceed to check whether the electroweak
vacuum is a global minimum, with the method introduced in Sec.\ \ref{sec:glo}.
For simplicity, we call it the GM (global minimum) condition. The
result is presented in Fig.\ \ref{fig:Constraints}, where the red
region is excluded by the BFB bound, the orange points violate the
GM conditions and the green points satisfy both the BFB and the GM
conditions. Two examples (one of them has been studied in Sec.\ \ref{sec:glo})
are marked in Fig.\ \ref{fig:Constraints} by black points, with
the numerical details given in Appendix \ref{sec:Numerical}, which
can be used to check the calculations.  The gray region represents
the constraint from direct searches from LHC, taken from \cite{Haber:2015pua}.
The yellow region violates the unitarity bound \cite{Casalbuoni:1986hy,Casalbuoni:1987eg,Maalampi:1991fb,Kanemura:1993hm,Arhrib:2000is,Akeroyd:2000wc,Ginzburg:2003fe,Ginzburg:2005dt,Horejsi:2005da},
checked by {\tt 2HDMC}. As is shown in the plot, there is a considerably
large part of the parameter space that can be excluded by the stability
of the electroweak vacuum at tree level. In the specific scenario
considered here, the GM constraint is even stronger than the LHC constraint
and complementary to the unitarity bound. One should note that this
is based on the type I $CP$-conserving 2HDM studied in Ref. \cite{Haber:2015pua}
in a particular scenario,  which should not be regarded as a general
conclusion. Nevertheless, the result presented in Fig.\ \ref{fig:Constraints}
suggests that  the vacuum stability at tree level  should be taken
into account in theoretical constraints on 2HDMs.

\section{Conclusion\label{sec:Conclusion}}

In the most general scalar potential of 2HDMs, more than one minimum
may coexist while the usually considered vacuum could be located at
a local minimum that could decay into a deeper one. We have seen such
an example in Sec.\ \ref{sec:glo}. To avoid vacuum instability at
tree level, we require a global minimum for the vacuum. Therefore
in this paper we study on the local and global minima of the 2HDM
potential and try to find out  the condition of a selected minimum
being globally minimal.

According to our analytical calculations, there are five different
possible types (denoted as type A to E) of minima, which are all summarized
in Tab.\ \ref{tab:All-local-minima}. Regarding the question of which
is the global minimum, though there has not been a simple answer,
a numerical process is proposed to address it, which is practically
applicable in phenomenological studies. 

The requirement of a global minimum will generate a new constraint
on the model. In a $CP$-conserving 2HDM with softly broken $\mathbb{Z}_{2}$
symmetry, we have shown in Fig.\ \ref{fig:Constraints} that such
a new constraint can be considerably significant in reducing the allowed
parameter space. In general, this constraint can be applied to many
other 2HDMs, which will be studied  in future work.

\begin{acknowledgments}
\textcolor{black}{XJX would like to thank Carlos Yaguna for some useful
discussions at the early stage of this work, and Werner Rodejohann
and Florian Goertz for carefully reading the manuscript and many helpful
suggestions. }
\end{acknowledgments}

\appendix

\section{Numerical examples\label{sec:Numerical}}

In this appendix we show two examples with numerical information in
detail. One example violates the GM condition  which will be called
example (1) below. The other satisfying the GM condition is called
example (2). Both are displayed in Fig.\ \ref{fig:Constraints}.

The input parameters in the hybrid basis are {[}here and henceforth
we add superscripts (1) and (2) on the corresponding quantities to
distinguish between them{]}
\begin{equation}
(m_{h},m_{H},\thinspace c_{\beta-\alpha},\thinspace Z_{4},\thinspace Z_{5},\thinspace Z_{7},\thinspace\tan\beta)^{(1)}=(125\thinspace{\rm GeV},500\thinspace{\rm GeV},\thinspace0.1,\thinspace-2,\thinspace-2,\thinspace0,\thinspace30),\label{eq:2h-46}
\end{equation}
\begin{equation}
(m_{h},m_{H},\thinspace c_{\beta-\alpha},\thinspace Z_{4},\thinspace Z_{5},\thinspace Z_{7},\thinspace\tan\beta)^{(2)}=(125\thinspace{\rm GeV},400\thinspace{\rm GeV},\thinspace0.1,\thinspace-2,\thinspace-2,\thinspace0,\thinspace10).\label{eq:2h-46-1}
\end{equation}

The output potentials computed by {\tt 2HDMC} are 
\begin{equation}
(m_{11}^{2},m_{22}^{2},m_{12}^{2})^{(1)}=(-0.110625,-0.00831996,0.00827196)\thinspace{\rm TeV}^{2},\label{eq:2h-30-1}
\end{equation}
\begin{equation}
\lambda_{1\cdots7}^{(1)}=(11.8234,0.270735,15.8106,-1.98716,-1.98716,-7.24247\times10^{-17},-8.88178\times10^{-16}),\label{eq:2h-31-1}
\end{equation}
\begin{equation}
(m_{11}^{2},m_{22}^{2},m_{12}^{2})^{(2)}=(0.0780633,-0.0062319,0.0158423)\thinspace{\rm TeV}^{2},\label{eq:2h-30-1-1}
\end{equation}
\begin{equation}
\lambda_{1\cdots7}^{(2)}=(2.62714,0.233919,6.60344,-1.97607,-1.97607,1.38778\times10^{-17},2.498\times10^{-16}).\label{eq:2h-31-1-1}
\end{equation}

The masses of scalar bosons computed according to Eqs.~(\ref{eq:2h-38}),
(\ref{eq:2h-38-1}), and (\ref{eq:2h-39}) are
\begin{equation}
(m_{h},\thinspace m_{H},\thinspace m_{A},\thinspace m_{H^{\pm}})^{(1)}=(125.0,500.0,607.376,607.376)\thinspace{\rm GeV},\label{eq:2h-47}
\end{equation}
\begin{equation}
(m_{h},\thinspace m_{H},\thinspace m_{A},\thinspace m_{H^{\pm}})^{(2)}=(125.0,400.0,528.967,528.967)\thinspace{\rm GeV}.\label{eq:2h-48}
\end{equation}
The mixing angle $\alpha$ from Eq.~(\ref{eq:2h-40}) is 
\begin{equation}
\alpha^{(1)}=0.0668463,\quad\alpha^{(2)}=0.000498351.\label{eq:2h-49}
\end{equation}

The first three types of minima computed according to Tab.\ \ref{tab:All-local-minima}
are

\begin{equation}
(V_{{\rm {\rm A}}},\thinspace V_{{\rm {\rm B}}},\thinspace V_{{\rm {\rm C}}})^{(1)}=(-0.333809,-1.2784,-5.17533)\cdot10^{-4}\thinspace{\rm TeV}^{4},\label{eq:2h-50}
\end{equation}
\begin{equation}
(q_{1},\thinspace q_{2},\thinspace z)_{{\rm {\rm A}}}^{(1)}=(0.411686,6.68905,-2.08135)\cdot10^{-3}\thinspace{\rm TeV}^{2},\label{eq:2h-51}
\end{equation}
\begin{equation}
(q_{1},\thinspace q_{2},\thinspace z)_{{\rm {\rm B}}}^{(1)}=(0,30.731,0)\cdot10^{-3}{\rm TeV}^{2},\label{eq:2h-52}
\end{equation}
\begin{equation}
(q_{1},\thinspace q_{2},\thinspace z)_{{\rm {\rm C}}}^{(1)}=(9.35648,0,0)\cdot10^{-3}{\rm TeV}^{2},\label{eq:2h-53}
\end{equation}
and 
\begin{equation}
(V_{{\rm {\rm A}}},\thinspace V_{{\rm {\rm B}}},\thinspace V_{{\rm C}})^{(2)}=(1.55994,-0.830128,-11.5979)\cdot10^{-4}\thinspace{\rm TeV}^{4},\label{eq:2h-54}
\end{equation}
\begin{equation}
(q_{1},\thinspace q_{2},\thinspace z)_{{\rm {\rm A}}}^{(2)}=(1.38198,-12.3714,-4.00854)\cdot10^{-3}\thinspace{\rm TeV}^{2},\label{eq:2h-55}
\end{equation}
\begin{equation}
(q_{1},\thinspace q_{2},\thinspace z)_{{\rm {\rm B}}}^{(2)}=(0,26.6413,0)\cdot10^{-3}\thinspace{\rm TeV}^{2},\label{eq:2h-56}
\end{equation}
\begin{equation}
(q_{1},\thinspace q_{2},\thinspace z)_{{\rm {\rm C}}}^{(2)}=(-29.7142,0,0)\cdot10^{-3}\thinspace{\rm TeV}^{2}.\label{eq:2h-57}
\end{equation}

The type D minima are obtained by numerically solving the non-linear
equations (\ref{eq:2h-23}), (\ref{eq:2h-24}) and (\ref{eq:2h-25}).
There are four different solutions in example (1) which are denoted
as D1, D2, D3 and D4:
\[
\left(\frac{q}{10^{-3}\thinspace{\rm TeV}^{2}},\thinspace\tan\beta,\thinspace\theta\right)_{{\rm D1}}^{(1)}=(9.41293,0.0807902,7.21702\times10^{-7}),
\]
\[
\left(\frac{q}{10^{-3}\thinspace{\rm TeV}^{2}},\thinspace\tan\beta,\thinspace\theta\right)_{{\rm D2}}^{(1)}=(30.3114,29.9987,6.28318),
\]
\[
\left(\frac{q}{10^{-3}\thinspace{\rm TeV}^{2}},\thinspace\tan\beta,\thinspace\theta\right)_{{\rm D3}}^{(1)}=(12.5001,4.51185,3.71351\times10^{-6}),
\]
\[
\left(\frac{q}{10^{-3}\thinspace{\rm TeV}^{2}},\thinspace\tan\beta,\thinspace\theta\right)_{{\rm D4}}^{(1)}=(6.5558,3.99351,3.14159).
\]
 The corresponding potential values are
\begin{equation}
(V_{{\rm D1}},\thinspace V_{{\rm D2}},\thinspace V_{{\rm D3}},\thinspace V_{{\rm D4}})^{(1)}=(-5.23783,-1.36168,-1.03783,-0.342805)\times10^{-4}\thinspace{\rm TeV}^{4}.\label{eq:2h-58}
\end{equation}
So D1 is the global minimum. 

As for example (2), there is only one solution at
\begin{equation}
\left(\frac{q}{10^{-3}\thinspace{\rm TeV}^{2}},\thinspace\tan\beta,\thinspace\theta\right)_{{\rm D}}^{(2)}=(30.3123,10.0,2.90367\times10^{-6}),\label{eq:2h-59}
\end{equation}
and the potential value is 
\begin{equation}
V_{{\rm D}}^{(2)}=-1.29348\times10^{-4}\thinspace{\rm TeV}^{4},\label{eq:2h-60}
\end{equation}
which is the global minimum of the potential in example (2).

\bibliographystyle{apsrev4-1}
\bibliography{ref}

\end{document}